\documentclass[twocolumn,showpacs,preprintnumbers,amsmath,amssymb,prl]{revtex4-2}
\usepackage{amsmath,amssymb}
\usepackage{graphicx}
\usepackage{booktabs}
\usepackage{xcolor}

\begin{document}

\title{Equations of Motion for an Economy:\\
Capital Deepening, Technology, and Firm Survival}

\author{Robert T.~Nachtrieb}
\affiliation{MIT Sloan School of Management, Cambridge, Massachusetts 02142, USA}

\begin{abstract}
We derive equations of motion for capital deepening in a competitive
economy directly from accounting identities, without assuming a
production function.
A profit imperative $\eta^* \equiv (w/\kappa + 1/\tau)/(1-f_p)$
sets the minimum viable capital productivity, where $\eta = Y/K$
[yr$^{-1}$] is capital productivity, $\kappa = K/L$ is capital
per worker, $w$ is the wage rate, $\tau$ is the capital lifetime,
and $f_p$ is the production tax share.
Four coupled relaxation equations govern $\kappa$, $\eta$,
the frontier productivity $\eta_{\rm new}$ of new investment,
and the labor share $q \equiv w/y$,
with the sandwich constraint $\eta^* \leq \eta_{\rm new} \leq \eta$
maintained as an exact invariant.
The frontier equation separates two physically distinct channels:
a structural cheapening channel ($\mu$, always active, drives
$\eta_{\rm new}$ downward) and a productivity channel ($\phi$,
historically zero).
Calibration against BEA 2-digit NAICS sector data (1998--2023)
confirms $\phi = 0$ for all identifiable sectors over 25 years;
the 75-year postwar record extends this finding across four
capital lifetimes.
A step $\phi = 0.01$\,yr$^{-1}$ --- a 1\%/yr improvement
in new-capital productivity, modest but historically unprecedented
--- nearly doubles the aggregate growth rate within one capital
lifetime, a falsifiable prediction with a precise observable
signature: upward-curving $\eta(t)$ in BEA sector data.
Firms near the zero-profit threshold have a cash martingale,
predicting establishment exit rate $\sim t^{-1/2}$; convolved
with the Zipf firm-size distribution~\cite{WP}, this yields
firm exit rate $\sim t^{-1/2}\!\log t$ with apparent exponent
$b = 0.295 \pm 0.03$, confirmed against BDS data with no free parameters.
\end{abstract}

\maketitle

\section{Introduction}

Macroeconomic growth theory typically begins with a production
function $Y = A_{\rm Solow}\,K^a L^{1-a}$ and treats total factor
productivity $A_{\rm Solow}$ as an unexplained residual~\cite{Solow}.
Here we take a different approach: we begin with accounting
identities that are exact at the firm level and derive equations
of motion for the observable intensive quantities of a sector ---
capital per worker $\kappa$, capital productivity $\eta \equiv Y/K$,
and the frontier productivity $\eta_{\rm new}$ of newly purchased
capital --- from first principles, without assuming a production
function.

The equations apply at the sector level, where firms serving the
same market with the same technology share approximately the same
intensive quantities ($\kappa$, $\eta$, $q \equiv w/y$), differing
only in scale.
Within-sector homogeneity implies that the distribution of intensive
quantities is tight; the ODEs govern their sector means, with the
calibrated RMSE of 5--10\% (Table~\ref{tab:calib}) placing an
empirical upper bound on the neglected within-sector dispersion.
A sharper bound is derived in~\cite{SM} by treating the Kalman filter
as a moment closure of the Fokker--Planck equation: the steady-state
filter covariance $P^*$, divided by the sector's Herfindahl--Hirschman
index, bounds the cross-firm variance of each intensive variable without
requiring micro data; labour share $q$ is confirmed scale-invariant
in every stable sector.
The full stochastic treatment (Fokker--Planck) is reserved for
firm-size and income distributions, derived in the companion
paper~\cite{WP}.

The result is a four-equation dynamical system with four
parameters per sector ($\beta$, $\gamma$, $\mu$, $\phi$)
identifiable from BEA national accounts.
A third parameter ($\phi$) governs a genuinely novel productivity
channel that has not appeared in 75 years of postwar data.
If and when it does, it will be unambiguous.

\section{Accounting framework and the profit imperative}

Starting from BEA national accounts~\cite{BEA}, value added per
worker is the exact identity:
\begin{equation}
  y \;\equiv\; \frac{Y}{L} \;=\; \eta\,\kappa.
  \label{eq:y}
\end{equation}
Free cash flow per unit capital is:
\begin{equation}
  \frac{\Pi}{K}
  \;=\; (1 - f_p)\bigl(\eta - \eta^*\bigr),
  \label{eq:Pi_K}
\end{equation}
where the \emph{profit imperative} is the breakeven capital
productivity at which $\Pi = 0$:
\begin{equation}
  \eta^*(\kappa,t) \;\equiv\;
  \frac{w(t)/\kappa \;+\; 1/\tau(t)}{1 - f_p(t)}.
  \label{eq:eta_star}
\end{equation}
All quantities are directly observable from BEA Fixed Assets and
GDP-by-Industry tables.
A sector with $\eta < \eta^*$ generates negative free cash flow
and cannot sustain investment.

\begin{figure*}[t]
  \includegraphics[width=\textwidth]{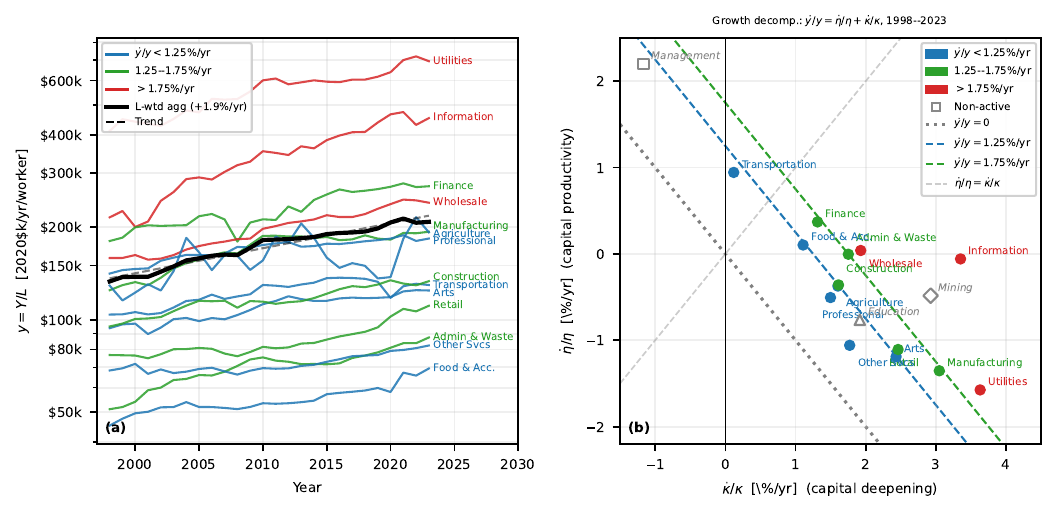}
  \caption{\textbf{Capital deepening drives output growth.}
    BEA 2-digit NAICS, 1998--2023 (2020\,\$), 14 active sectors,
    L-weighted; Real Estate excluded (imputed rent artefact).
    \textit{(a)} Real output per worker $y = \eta\kappa$ by sector
    (log scale); colours by growth rate: blue $<1.25\%$\,yr$^{-1}$,
    green $1.25$--$1.75\%$, crimson $>1.75\%$;
    black = L-weighted aggregate ($+1.9\%$\,yr$^{-1}$).
    \textit{(b)} Growth decomposition
    $\dot y/y = \dot\eta/\eta + \dot\kappa/\kappa$;
    open markers: non-active sectors.
    All sectors lie above $\dot y/y = 0$; capital deepening dominates;
    $\dot\eta/\eta \leq 0$ in 13 of 17 sectors.}
  \label{fig:y_real}
\end{figure*}

Figure~\ref{fig:y_real} shows the key empirical facts.
Real output per worker $y$ (active sectors, L-weighted) grows
at $+1.9\%$\,yr$^{-1}$ over 1998--2023.
Capital deepening --- not rising $\eta$ --- accounts for
essentially all of this growth: $\dot\kappa/\kappa > 0$ in
every sector, while $\dot\eta/\eta \leq 0$ in 13 of 17 sectors.
The identity $\dot y/y = \dot\eta/\eta + \dot\kappa/\kappa$
shows that $\eta$ is not technology --- it is the
capital-output ratio reciprocal, an accounting quantity that
falls as investment deepens.


\section{Four coupled equations of motion}

The state variables are $\kappa$, $\eta$, $\eta_{\rm new}$, and
the labor share $q \equiv w/y$, with $w = q\,\eta\,\kappa$ closing
the model.
Since new capital is more capital-intensive than the installed base,
$\eta_{\rm new} \leq \eta$; since no rational firm invests below the
profit threshold, $\eta_{\rm new} \geq \eta^*$.
The \emph{sandwich constraint}
\begin{equation}
  \eta^*(t) \;\leq\; \eta_{\rm new}(t) \;\leq\; \eta(t)
  \label{eq:sandwich}
\end{equation}
is preserved as an exact invariant of the ODE system~\cite{SM}.
The four equations are:
\begin{align}
  \frac{1}{\kappa}\frac{d\kappa}{dt}
  &\;=\; \beta\,(1-f_p)\,(\eta - \eta^*),
  \label{eq:kappa_dot} \\[3pt]
  \frac{dq}{dt}
  &\;=\; \gamma\,(q_{\rm asy} - q),
  \qquad 0 \leq q_{\rm asy} \leq 1,
  \label{eq:q_dot} \\[3pt]
  \frac{d\eta}{dt}
  &\;=\; (\eta_{\rm new} - \eta)
         \!\left(\frac{1}{\tau} + g\right),
  \label{eq:eta_dot} \\[3pt]
  \frac{d\eta_{\rm new}}{dt}
  &\;=\; \underbrace{\mu\,(\eta^* - \eta_{\rm new})}_{\text{structural}}
       + \underbrace{\phi\,\eta_{\rm new}}_{\text{productivity}}.
  \label{eq:eta_new_dot}
\end{align}
Here $g = \beta(1-f_p)(\eta-\eta^*)$ is the net investment rate,
and Eq.~\eqref{eq:eta_dot} is derived from a co-flow accounting
identity for technology-endowed capital~\cite{SM}.

Equations~\eqref{eq:kappa_dot}--\eqref{eq:eta_dot} share the same
relaxation-toward-a-moving-target structure.
The labor share $q$ relaxes to sector-specific $q_{\rm asy}$
(Kaldor's stylized fact~\cite{Kaldor1957}), closing the wage
channel without an exogenous wage series.
Equation~\eqref{eq:eta_new_dot} is the key structural result:
it separates two physically distinct mechanisms.
The \textbf{structural channel} $\mu$ [yr$^{-1}$] describes
diminishing returns within a sector: each successive generation of
capital requires greater investment per unit output
($\eta_{\rm new} = \Delta Y/\Delta K < \eta$, so $\eta_{\rm new}$
falls toward $\eta^*$, the minimum viable threshold).
Nobody decides $\mu$ and nobody wants $\eta_{\rm new}$ to fall,
but it does --- relentlessly, in every postwar technological wave.
The \textbf{productivity channel} $\phi$ [yr$^{-1}$] would
raise $\eta_{\rm new}$ if new capital were genuinely more
capable per dollar than the current frontier.
Crucially, $\partial(\dot y/y)/\partial\mu < 0$ at $\phi = 0$:
faster cheapening \emph{reduces} medium-run growth by compressing
profit margins faster than capital deepening can compensate.

\section{Calibration and the productivity channel}

Equations~\eqref{eq:kappa_dot}--\eqref{eq:eta_new_dot} are
calibrated sector-by-sector to BEA 2-digit NAICS data (1998--2023),
fitting $\kappa(t)$ and $\eta(t)$ simultaneously via Nelder-Mead
in log-parameter space, with $w(t)$, $\tau(t)$, $f_p(t)$
interpolated from the BEA panel.
Table~\ref{tab:calib} summarises results.
$\beta$ spans a factor of $\sim 30$ (Constr $0.02$ to Util $0.88$),
reflecting genuine heterogeneity in investment responsiveness.
Five sectors have $\mu \to 0$ (pipeline stationary on a
25-year timescale); two sectors (OtherSvc, Retail) yield large
$\mu$ that absorb high-frequency noise and should be treated
with caution.\footnote{Sectors with $\mu \to 0$ may alternatively
reflect the balanced-frontier regime ($0 < \phi < \mu$, fixed point
$\eta_{\rm new}^* = \mu\eta^*/(\mu-\phi)$), which is observationally
degenerate with $\mu = 0$ over a 25-year window.
See~\cite{SM} for the Kalman filter analysis distinguishing the two regimes.}

The 95\% confidence interval for $\phi$ includes zero for every
sector (Table~\ref{tab:calib}).
Six sectors are \emph{identifiably} $\phi = 0$ (upper CI
$< 0.15$\,yr$^{-1}$): Manufacturing, Utilities, Arts,
Retail, Professional Services, Construction.
The remaining eight have wide CIs --- $\phi$ is unidentifiable
from 25 years of data, not evidence of $\phi > 0$.
The 75-year extended panel (SIC-basis NIPA 1948--1987 spliced
to NAICS 1998--2023) confirms $\phi = 0$ across four capital
lifetimes in Manufacturing and Finance: every postwar technological
wave --- electrification, computers, the internet --- operated
through the structural channel $\mu$, not the productivity
channel $\phi$~\cite{SM}.

\begin{table}[h]
\caption{Calibrated parameters, 14 active sectors
  (BEA NAICS, 1998--2023).
  Each sector: best estimate (row~1) and
  95\% CI in brackets (row~2).
  $\beta$ (dimensionless): investment responsiveness.
  $\mu$ [yr$^{-1}$]: pipeline rate;
  $\to\!0$ means $\mu<10^{-3}$\,yr$^{-1}$.
  $\phi$ [yr$^{-1}$]: all CIs include zero;
  upper bound $10.00$ is the scan boundary, not a posterior bound.
  Full table in~\cite{SM}.}
\label{tab:calib}
\smallskip
\setlength{\tabcolsep}{3pt}
\scriptsize
\begin{tabular}{@{}lp{1.6cm}p{1.6cm}p{1.6cm}@{}}
\toprule
Sector & $\beta$ & $\mu$ & $\phi$ \\
\midrule
Utilities & $0.876$ & $0.154$ & $0.000$ \\
& $[0.849,\,0.902]$ & $[0.108,\,0.222]$ & $[0.000,\,0.052]$ \\[2pt]
Information & $0.236$ & $0.001$ & $0.000$ \\
& $[0.222,\,0.249]$ & $[0.000,\,0.009]$ & $[0.000,\,2.454]$ \\[2pt]
Arts & $0.220$ & $0.097$ & $0.000$ \\
& $[0.185,\,0.252]$ & $[0.041,\,0.217]$ & $[0.000,\,0.131]$ \\[2pt]
Manuf. & $0.185$ & $0.089$ & $0.000$ \\
& $[0.171,\,0.199]$ & $[0.069,\,0.109]$ & $[0.000,\,0.052]$ \\[2pt]
Other Svcs & $0.167$ & $\to 0$ & $0.000$ \\
& $[0.103,\,0.175]$ &  & $[0.000,\,4.406]$ \\[2pt]
Retail & $0.125$ & $0.307$ & $0.013$ \\
& $[0.109,\,0.140]$ & $[0.104,\,2.000]$ & $[0.000,\,0.104]$ \\[2pt]
Agriculture & $0.108$ & $0.010$ & $0.000$ \\
& $[0.098,\,0.118]$ & $[0.000,\,0.038]$ & $[0.000,\,0.855]$ \\[2pt]
Food \& Acc. & $0.080$ & $\to 0$ & $2.183$ \\
& $[0.059,\,0.101]$ &  & $[0.000,\,10.00]$ \\[2pt]
Admin & $0.050$ & $\to 0$ & $0.677$ \\
& $[0.043,\,0.057]$ &  & $[0.000,\,3.101]$ \\[2pt]
Finance & $0.043$ & $\to 0$ & $2.759$ \\
& $[0.038,\,0.048]$ &  & $[0.000,\,7.912]$ \\[2pt]
Transport & $0.010$ & $\to 0$ & $7.912$ \\
& $[0.000,\,0.038]$ &  & $[0.000,\,10.00]$ \\[2pt]
Prof.\ Svcs & $0.036$ & $0.027$ & $0.000$ \\
& $[0.030,\,0.041]$ & $[0.010,\,0.045]$ & $[0.000,\,0.148]$ \\[2pt]
Wholesale & $0.040$ & $\to 0$ & $0.962$ \\
& $[0.036,\,0.043]$ &  & $[0.000,\,4.406]$ \\[2pt]
Constr. & $0.019$ & $0.012$ & $0.000$ \\
& $[0.013,\,0.025]$ & $[0.000,\,0.055]$ & $[0.000,\,0.677]$ \\[2pt]
\bottomrule
\end{tabular}
\end{table}

\begin{figure}[t]
  \includegraphics[width=\columnwidth]{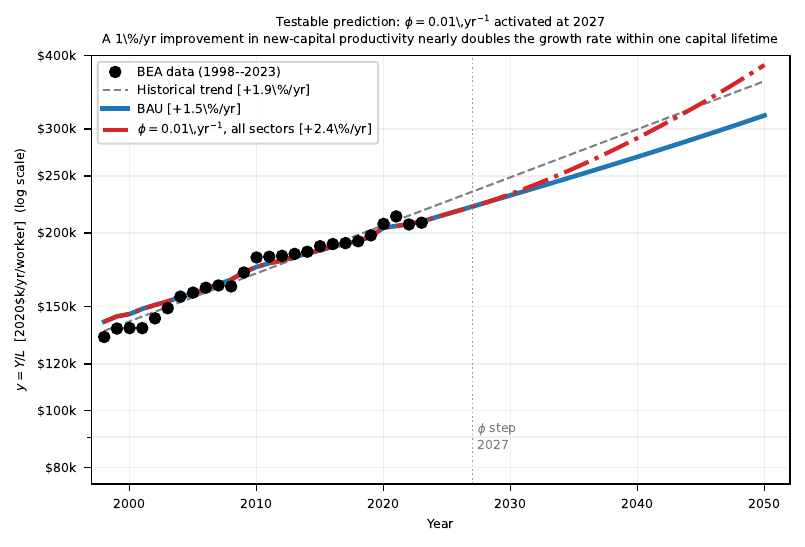}
  \caption{\textbf{Falsifiable prediction: $\phi = 0.01$\,yr$^{-1}$
    at 2027, all 14 active sectors.}
    Black dots: BEA observed $y(t)$, 1998--2023.
    Grey dashed: historical trend ($+1.9\%$\,yr$^{-1}$).
    Blue: business-as-usual ($+1.5\%$\,yr$^{-1}$).
    Red dash-dot: $\phi = 0.01$\,yr$^{-1}$ ($+2.4\%$\,yr$^{-1}$).
    A 1\%/yr improvement in new-capital productivity nearly doubles
    the growth rate within one capital lifetime ($\tau \approx 13$\,yr).
    Observable signature: upward-curving $\eta(t)$ in BEA data,
    absent from 75 years of postwar history.}
  \label{fig:whatif}
\end{figure}

\paragraph{Falsifiable prediction.}
If AI-embodied capital represents a genuine productivity
improvement --- not merely cheaper capital, but more capable
capital per dollar --- the parameter $\phi$ becomes positive
for the first time in the postwar record.
A step $\phi = 0.01$\,yr$^{-1}$ beginning in 2027,
applied to all 14 active sectors, nearly doubles the
aggregate growth rate: from $+1.5\%$ to $+2.4\%$\,yr$^{-1}$
by 2050 (Fig.~\ref{fig:whatif}).
This is a 1\%/yr improvement in new-capital productivity ---
modest in absolute terms.
At this rate $\eta_{\rm new}$ grows by a factor of
$e^{0.01 \times 23} \approx 1.26$ between 2027 and 2050,
well within the stable regime.
The observable signature is upward-curving $\eta(t)$ in
BEA annual data --- something absent from 75 years of
postwar history~\cite{SM}.
If it appears, it will be unambiguous.

\section{Connection to firm survival}

The profit imperative connects macro dynamics to the micro
firm-size statistics of Ref.~\cite{WP}.
Firms near the lower margin --- where $\eta_i \approx \eta^*$
--- have $\Pi/K \approx 0$.
Their cash reserve evolves as:
\begin{equation}
  C(t) = C(0) + \int_0^t \!\bigl[(1-f_c)\,\Pi(t') - I(t')\bigr]\,dt',
  \label{eq:cash}
\end{equation}
where $f_c$ is the corporate tax rate and $I(t')$ is investment.
At $\eta \approx \eta^*$, $(1-f_c)\Pi \approx I$, so $C(t)$ has
zero conditional expected change under i.i.d.\ cash-flow shocks:
it is a martingale~\cite{SM}.
The first-passage time to $C = 0$ gives establishment exit
rate $S_e(t) \sim t^{-1/2}$, confirmed in BDS establishment-age
data~\cite{BDS}.

A firm with $n$ establishments fails when the last exits:
$T_{\rm firm} = \max(T_1,\ldots,T_n)$.
Averaging over the Pareto firm-size distribution
$p(n) \sim n^{-(\alpha+1)}$ (derived in Ref.~\cite{WP})
via the Laplace transform:
\begin{equation}
  S_{\rm firm}(t) \;=\; 1 - \mathcal{L}_n\!\bigl(A\,t^{-1/2}\bigr).
  \label{eq:Sfirm}
\end{equation}
At $\alpha = 1$ (Zipf), $\mathcal{L}_n(s) \sim 1 - cs\log(1/s)$,
giving:
\begin{equation}
  \boxed{S_{\rm firm}(t) \;\sim\;
  \tfrac{cA}{2}\;t^{-1/2}\log t \quad (\alpha = 1).}
  \label{eq:Sfirm_Zipf}
\end{equation}
A power-law fit over $t = 1$--$30$\,yr yields apparent
exponent $b_{\rm app} \approx 0.32$ (analytic formula);
the direct fit to BDS age bins gives $b = 0.295 \pm 0.03$,
consistent within $1\sigma$; Monte Carlo simulation at $\alpha=1$
gives $b_{\rm sim} \approx 0.335$~\cite{SM}.
No parameters are tuned: $\alpha$ from CBP~\cite{CBP},
$b$ from BDS~\cite{BDS}.


\section{Discussion}

The four coupled equations~\eqref{eq:kappa_dot}--\eqref{eq:eta_new_dot}
constitute a minimal dynamical model of a competitive sector,
derived entirely from accounting identities.
No production function, no equilibrium assumption, and no
exogenous technology process are required.

The central empirical finding is that $\phi = 0$ across the
full postwar record.
Every technological wave observed in the data --- electrification,
automation, computers, the internet --- operated through the
structural channel $\mu$: capital became cheaper per unit output,
sustaining investment through capital deepening, but did not
become more capable per dollar.
This is not a modest claim: it means that no prior technology shock
has shifted the economy onto a genuinely more productive growth
trajectory in the sense captured by $\phi > 0$.

The counterintuitive policy implication follows directly:
raising $\mu$ --- accelerating the cheapening pipeline ---
\emph{reduces} medium-run growth at $\phi = 0$, because faster
cheapening compresses profit margins faster than capital deepening
can compensate.
The correct lever is $\phi$, not $\mu$.
Whether AI represents the first historically observed instance
of $\phi > 0$ is an empirical question, answerable from BEA
annual data within one capital lifetime of AI's large-scale
deployment.

The framework connects to the companion paper~\cite{WP} at two
points.
First, within-sector homogeneity --- firms in a sector share
approximately the same $\kappa$, $\eta$, and $q$, differing only
in scale --- justifies the ODE treatment of sector means, while
the full distributional dynamics (Fokker--Planck) govern the
firm-size tails treated in Ref.~\cite{WP}.
Second, the profit imperative $\eta^*$ is the same quantity
that determines whether a firm's cash process is a martingale,
linking the macro ODE system to the micro exit-rate prediction.
Macro drift and micro noise are complementary descriptions of
the same economy.

\begin{acknowledgments}
The author thanks D.~Peterson for the insight connecting capital
productivity to embedded technology, and V.~Yakovenko for discussions
on the relationship between firm dynamics and macroeconomic
observables.
\end{acknowledgments}


\begin{thebibliography}{99}
\bibitem{SM} R.~Nachtrieb, Supplemental Material for this article,
  including BEA/BDS/CBP data pipelines, derivations, and
  sector-by-sector calibration figures (2026).
\bibitem{Solow} R.~M.~Solow, Rev.\ Econ.\ Stat.\ \textbf{39}, 312 (1957).
\bibitem{WP} R.~Nachtrieb, \textit{Boltzmann-Gibbs Income Distributions
  and Zipf Firm Sizes from Non-Equilibrium Statistical Mechanics}
  (companion paper, 2026).
\bibitem{BEA} U.S.\ Bureau of Economic Analysis, GDP by Industry
  and Fixed Assets accounts, \texttt{apps.bea.gov} (2024).
\bibitem{BDS} U.S.\ Census Bureau, Business Dynamics Statistics,
  Firm Age series 1978--2023, \texttt{census.gov/programs-surveys/bds}
  (2024).
\bibitem{CBP} U.S.\ Census Bureau, County Business Patterns,
  2017--2023, \texttt{census.gov/programs-surveys/cbp} (2024).
\bibitem{Kaldor1957} N.\ Kaldor,
  \textit{A Model of Economic Growth},
  Econ.\ J.\ \textbf{67}(268), 591--624 (1957).
\bibitem{Maddison} M.~Bolt \textit{et al.}, Maddison Project Database
  2023, \texttt{ggdc.net/maddison} (2023).
\bibitem{Axtell} R.~L.~Axtell, Science \textbf{293}, 1818 (2001).
\end{thebibliography}
\end{document}